\documentclass[aps,pre,amsmath,amssymb,twocolumn,showpacs,byrevtex,superscriptaddress]{revtex4}
\newcommand{\on}[2]{\mathop{\null#2}\limits^{#1}} 
\newcommand{\oover}[1]{\on{\circ}{#1}} 
\usepackage{graphicx}
\begin{document}
\title{Scaling of dynamics with the range of
interaction  in short-range attractive colloids}.

\author{G.~Foffi}\affiliation{
         {Dipartimento di Fisica and INFM,
         Universit\`a di Roma {\em La Sapienza}, P.le A. Moro 2, 00185 Roma, Italy}
        }\affiliation{Institut Romand de Recherche Num\'erique en Physique des Mat\'eriaux IRRMA, PPH-Ecublens, CH-105 Lausanne, Suisse}

\author{C.~De Michele}\affiliation{
         {Dipartimento di Fisica and INFM,
         Universit\`a di Roma {\em La Sapienza}, P.le A. Moro 2, 00185 Roma, Italy}
        }\affiliation{
        {INFM - CRS Soft,  Universit\`a di Roma {\em La Sapienza}, P.le A. Moro 2, 00185 Roma, Italy}
        }

\author{F.~Sciortino} \affiliation{
         {Dipartimento di Fisica and INFM,
         Universit\`a di Roma {\em La Sapienza}, P.le A. Moro 2, 00185 Roma, Italy}
        }\affiliation{
        {INFM - CRS Soft,  Universit\`a di Roma {\em La Sapienza}, P.le A. Moro 2, 00185 Roma, Italy}
        }
\author{P.~Tartaglia} \affiliation{
         {Dipartimento di Fisica and INFM,
         Universit\`a di Roma {\em La Sapienza}, P.le A. Moro 2, 00185 Roma, Italy}
        }\affiliation{
         {INFM - CRS SMC,  Universit\`a di Roma {\em La Sapienza}, P.le A. Moro 2, 00185 Roma, Italy}
}
\begin{abstract}
We numerically study the dependence of the dynamics on the range of
interaction $\Delta$ for the short-range square well potential. We
find that, for small $\Delta$, dynamics scale exactly in the same way
as thermodynamics, both for Newtonian and Brownian microscopic
dynamics.  For interaction ranges from a few percent down to the
Baxter limit, the relative location of the attractive glass line and
the liquid-gas line does not depend on $\Delta$.  This proves that in
this class of potentials, disordered arrested states (gels) can be
generated only as a result of a kinetically arrested phase separation.
\end{abstract}


\maketitle

\begin{figure}[!h]
\includegraphics[width=.5\textwidth]{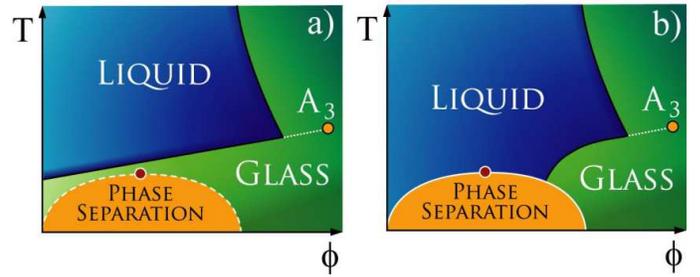}
\caption{Sketch of two possible relative location in the $T$-$\phi$ 
plane of the liquid-gas coexistence line and of the glass lines. In a)
the liquid-gas coexistence is hidden below the liquid-glass transition
line (case (i) in the text). In b) the glass line intersect the
binodal below the critical temperature at a $\phi$ value higher then
the critical one (case (ii) in the text). The $A_3$ point is the MCT
singularity
\protect~\cite{Fabbian1999etal,Bergenholtz1999,Dawson2001etal} that might be encountered
for the range values we discuss in this work.}
\label{fig:sketch}
\end{figure}

Colloidal dispersions form gels, disordered arrested states of matter
at low packing fraction $\phi$, if the colloid-colloid hard sphere
repulsion is complemented by a short-range
attraction~\cite{Poon1998,Verhaegh1999etal,Segre2001etal,Shah2003etal}.
The nature of the gel transition in short-range attractive colloidal
systems has received significant attention in recent years (for a
recent review see for example Ref.~\cite{Trappe2004}). Several
routes to the gel state have been proposed and critically examined.
In particular,
it has been speculated that the gel-line constitutes the extension to
low $\phi$ of the attractive-glass line, an idea which would provide
an unifying interpretation of the gel and glass arrest state of
matter~\cite{Bergenholtz2003etal}.  An alternative interpretation
suggests that colloidal gel results from an interrupted liquid-gas
phase separation, interrupted by the glass transition which takes
place in the dense regions created during the spinodal decomposition
kinetics~\cite{Soga1998etal,Lodge1999,Zaccarelli2004etal}. The two scenarios,
which differ only by the relative location of the glass line(s)
with respect to the phase separation line, are sketched in
Fig.\ref{fig:sketch}.  In case (i), the attractive glass line pre-empts
the meta-stable liquid-gas separation and the gel line can be
approached from equilibrium conditions (Fig.\ref{fig:sketch}a).  In
case (ii), the glass line meets the phase separation line on the high
$\phi$ branch, and the morphology of the low $\phi$ arrested state is
dictated by the phase separation kinetic (Fig.\ref{fig:sketch}b).
\\
The thermodynamic phase diagram of simple models for short-range
attraction has been evaluated theoretically and successfully compared
with experimental
data~\cite{Lekkerkerker1992etal,Tejero1994etal,Hagen1994,Dijkstra1999etal}.  When
the attraction range is a few percent of the particle size, the
equilibrium phase diagram is composed only by a fluid-phase and a
crystalline phase. The liquid-gas coexistence locus is hidden within
the region of fluid-crystal coexistence. For small range of
attractions, the liquid-gas coexistence curve for different models can
be scaled onto each other by comparing different systems at the same
value of the second virial coefficient~\cite{Noro2000}, providing an
effective characterization of the dependence of the liquid-gas
coexistence line on the range of attraction.

The dependence of dynamic properties on the range of attraction has
been studied at large $\phi$ within the mode-coupling
theory~\cite{Dawson2001etal}. For small ranges, two distinct glass
lines appear, indicating the possibility of forming two distinct glass
states, commonly named repulsive and attractive
glasses~\cite{Fabbian1999etal,Bergenholtz1999,Dawson2001etal,Pham2002etal,Eckert2002,Foffi2004cetal}. According
to
MCT~\cite{Fabbian1999etal,Bergenholtz1999,Dawson2001etal,Foffi2002etal},
for small attraction range, the attractive glass line extends to low
$\phi$, supporting type (i) scenario.  Unfortunately, MCT
overestimates the location of the glass lines and a non negligible
mapping must be applied on theoretical curves before comparing theory
with experimental or simulation
data~\cite{Sperl2003b,Sciortino2003etal}.  In the case of a square well
potential with an attractive range of 3\% of the hard-sphere diameter
such a mapping has been evaluated and the mapped attractive glass line
has been found to end on the right side of the spinodal~\cite{
Zaccarelli2004etal} , in agreement with type (ii) scenario.\\
If the location of the attractive glass line and of the liquid-gas
line depends in different ways on the range of the attraction a
transition from case (ii) (which is known to be the correct case for
interaction ranges larger than 3\%~\cite{
Zaccarelli2004etal} ) to case (i) could take place at a
very small critical value of the attraction range. This Letter
addresses this question, by examining the dependence of the dynamics
on the attraction range, both for Brownian (BD) and Newtonian (ND)
dynamics.  We show that, for interaction ranges
from a few percent down to the Baxter limit~\cite{Baxter1968b} and for packing fraction smaller than $0.40$,
(the packing fraction range where gels are observed in experiments),
dynamics and thermodynamics loci scale with the range of
interaction in the same way, ruling out case (i) as route to the gel
formation in short-range attractive colloids.

We investigate a system that has been extensively studied earlier, a
binary square well (SW) mixture~\cite{Zaccarelli2002betal,Foffi2004betal}.
The binary system is a $50\%$-$50\%$ mixture of $N=2000$
particles. The two species (labeled $A$ and $B$) are characterized by
a diameter ratio $\sigma_A/\sigma_B=1.2$. Masses are chosen to be
equal and unitary. The attraction is modeled by a SW interaction
defined according to:
\begin{equation}
\label{pote}
V^{\alpha,\beta}(r) = \left\{
  \begin{array}{ll}
    \infty & \mbox{ $r<\sigma_{\alpha,\beta}$}
\\
    -u_0 & \mbox{$\sigma_{\alpha,\beta}<r<\sigma_{\alpha,\beta}+\Delta_{\alpha,\beta}$}
\\
    0 & \mbox{$r>\sigma_{\alpha,\beta}+\Delta_{\alpha,\beta}$}
  \end{array} \right.
\end{equation}

where $\sigma_{\alpha,\beta}=(\sigma_\alpha+\sigma_\beta)/2$,
$\alpha,\beta=A,B$ and $\Delta_{\alpha,\beta}$ is the range of the
attraction.  We fix $\sigma_{\alpha,\beta}$ and vary the relative
well-width $\epsilon \equiv
\frac{\Delta_{\alpha,\beta}}{\Delta_{\alpha,\beta}+\sigma_{\alpha,\beta}}$.  
We report data for extremely small relative well width ---
from $10^{-2}$ to $5 \cdot 10^{-6}$ --- covering an interval
which starts from a physically realizable limit and ends close  to the
theoretical Baxter limit.
We choose $k_B=1$ and set the depth of the potential $u_0=1$. Hence
$T=1$ corresponds to a thermal energy $k_BT$ equal to the attractive
well depth. The diameter of the small specie is chosen as unity of
length, i.e. $\sigma_B=1$. Density is expressed in term of packing
fraction $\phi=(\rho_A \sigma_A^3+\rho_B \sigma_B^3)\cdot \pi/6$,
where $\rho_\alpha=N_\alpha/L^3$, $L$ being the box size and
$N_\alpha$ the number of particles of species $\alpha$. Time is
measured in units of $\sigma_B\cdot(m/u_0)^{1/2}$.  ND has been coded via a standard  event driven
algorithm, commonly used  for particles interacting with
step-wise potentials~\cite{Rapaport97}. 
BD has
been implemented via the position Langevin equation:
\begin{equation}
\dot {\bold r_i} (t) = \frac{D_0}{k_B T} {\bf f}_i(t) + {\oover{\bf r}}_i(t), 
\label{Eq:langevin}
\end{equation}
\noindent
coding the algorithm developed by Strating~\cite{Strating1999}.  In
Eq.\ref{Eq:langevin} $ {\bold r_i} (t) $ is the position of particle
$i$, ${\bf f}_i(t)$ is the total force acting on the particle, $D_0$
is the short-time (bare) diffusion coefficient, $ {\oover{\bf r}}_i(t)
$ a random thermal noise satisfying $< {\oover{\bf r}}_i(t)
{\oover{\bf r}}_i(0)> = k_BT \delta(t)$.  In Strating's algorithm, a
random velocity (extracted from a Gaussian distribution of variance
$\sqrt{k_BT/m}$) is assigned to each particle and the system is
propagated for a finite time-step $\frac{2 mD_0}{k_BT}$, according to
event-driven dynamics. We chose $D_0$ such that short time motion is
diffusive over distances smaller than the well width.  For the
smallest $\epsilon$, reliable estimates of dynamical properties
require more than 10$^{10}$ collisions (about two weeks on a 3GHz processor).

For the very small values of relative well width $\epsilon$ considered
here, thermodynamic properties at different well width can be scaled
by using as scaling variable the value of the second virial
coefficient $B_2 $~\cite{Noro2000}.
For the SW binary mixture
$B_2=\frac{B_2^{AA}+B_2^{BB}+2B_2^{AB}}{4} $ where
\begin{equation}
B_2^{\alpha,\beta}=\frac{2}{3} \pi \sigma^3_{\alpha,\beta}   \left[1 - (e^{\beta u_0}-1)[(1-\epsilon)^{-3}-1]\right].
\end{equation}

For the equivalent 50-50 hard-sphere binary mixture, $B_2$ is
\begin{equation}
B_2^{HS}=\frac{2}{3} \pi \frac{[ \sigma^3_{AA}+
\sigma^3_{BB}+2 \sigma^3_{AB}]}{4}
\end{equation}
An adimensional second virial coefficient can be defined as $B_2^*
\equiv B_2/B_2^{HS}$. This quantity helps in comparing between
different models and different samples~\cite{Rosenbaum1996etal}.  At
small $\epsilon$, $B_2^*$ becomes essentially
function of the variable $\epsilon e^{\beta u_0} \approx \Delta
e^{\beta u_0}$ .  In the same limit, state points at the same $B_2^*$
and $\phi$ are characterized, to a very good approximation, by same
thermodynamic properties, i.e. same bonding pattern, same energy, same
structure. In the limit $\epsilon \rightarrow 0$ the system behaves
similarly to the Baxter model~\cite{Baxter1968b} at the same $B_2^*$
state point.  The Baxter potential $V_B(r)$ is best defined via
\begin{equation}
\label{bax-po}
e^{-\beta V_B(r)} = \theta(r-\sigma) +\frac{\sigma}{12 \tau} \delta(r-\sigma)
\end{equation}
where $\tau$ is the adhesiveness parameter, which plays the role of
effective temperature, $\theta$ and $\delta$ are respectively the
Heaviside and Dirac functions. This model has been extensively used in
the interpretation of experimental data~\cite{Chen1994etal} despite its known
pathologies~\cite{Fishman1981,Stell1991}.  For the Baxter potential,
$B_2^*=1-1/4\tau$ and the location of the liquid-gas critical point,
recently determined with great accuracy, is $\phi_c=0.266
$ and $B_{2}^*=-1.2$~\cite{Miller2003}.

\begin{figure}[th!]
\includegraphics[width=.4\textwidth]{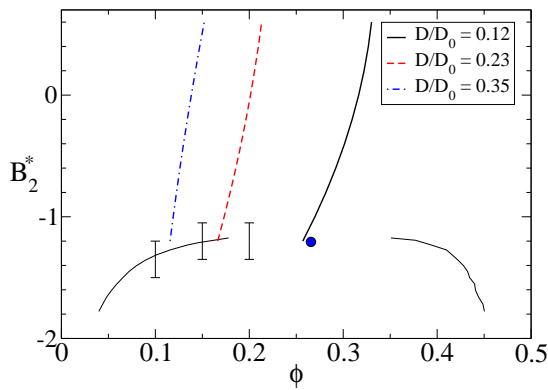}
\caption{Phase diagram for $\epsilon=0.01$. The continuous line reproduces
the coexistence curve calculated  by Miller and Frenkel
\protect~\cite{Miller2003}. 
The error bars represent our results, i.e. the intervals which bracket
the spinodal (see text). Isodiffusivity lines for three typical liquid
values of $D/D_0$ are plotted. Note that slow dynamics pre-glassy
features (like two-step relaxation decay) appear only when $D/D_0
\lesssim 5\cdot10^{-4}$.}
\label{fig:1}
\end{figure}

Fig.~\ref{fig:1} shows the spinodal line for the SW model with
$\epsilon =0.01$, estimated by bracketing it with the lowest $T$
stable point and the first phase separating state point along each
isochore.  It also show the data from Miller and
Frenkel~\cite{Miller2003} for the Baxter potential. The agreement
between the two set of data, notwithstanding the different system
(binary mixture vs. one component, SW vs. Baxter) confirms that the
Baxter limit is already reached when $\epsilon=0.01$. Fig.~\ref{fig:1}
also shows the ND isodiffusivity lines~\cite{Foffi2002cetal}, defined as
the locus where the normalized diffusion coefficient $D/D_0$ is
constant.  For ND, the normalization factor $D_0 \equiv \sqrt{ 3 k_BT
\sigma^2/m}=v_{th} \sigma_{B}$ accounts for differences in the
microscopic time due to different thermal velocity $v_{th}$.  The
large values of $D/D_0$, even for $\phi \approx \phi_c$, confirms
that, as in the previously studied $\epsilon=0.03$
case~\cite{Zaccarelli2004etal}, no arrested states can be approached in
equilibrium for $\phi < \phi_c$.

We next address the question of the dependence of the dynamics on
$\epsilon$.  We focus on two specific values of $\phi$, respectively
on on the left ($\phi=0.2$) and on the right ($\phi=0.4$) of the
critical point.  For each $\phi$, we select several pairs of
$\epsilon-T$ values such that $B_2^*=-0.405$. The average energy per
particle are respectively of $-2$ and $-4$.  Within our numerical
precision, simulations for different $\epsilon-T$ values converge to
the same average potential energy and same structure, supporting the
hypothesis that for these small $\epsilon$ values equality in $B_2^*$
implies equal thermodynamic properties.

We focus on two dynamic quantities, the tagged-particle mean square
displacement $<r^2(t)>$ and the bond autocorrelation function $C(t)$,
defined by:
\begin{equation}
\label{corre_bo}
C(t)=\langle \sum_{i<j}^{1,N} c_{ij}(0)c_{ij}(t)\rangle / {\langle  \sum_{i<j}^{1,N} c_{ij}^2(0)\rangle}
\end{equation}
where the $N\times N$ matrix $c_{ij}(t)$ defines the bonds at time $t$ according to:
 \begin{equation}
\label{bo_matr}
c_{ij}(t) = \left\{
  \begin{array}{ll}
    1 & \mbox{  if $i$ and $j$ particles are bonded}
 \\
    0 & \mbox{  else}
 \end{array} \right.
\end{equation}
\noindent
Two particles are considered bonded if their relative distance is in the attractive well. 

\begin{figure}[tbh]
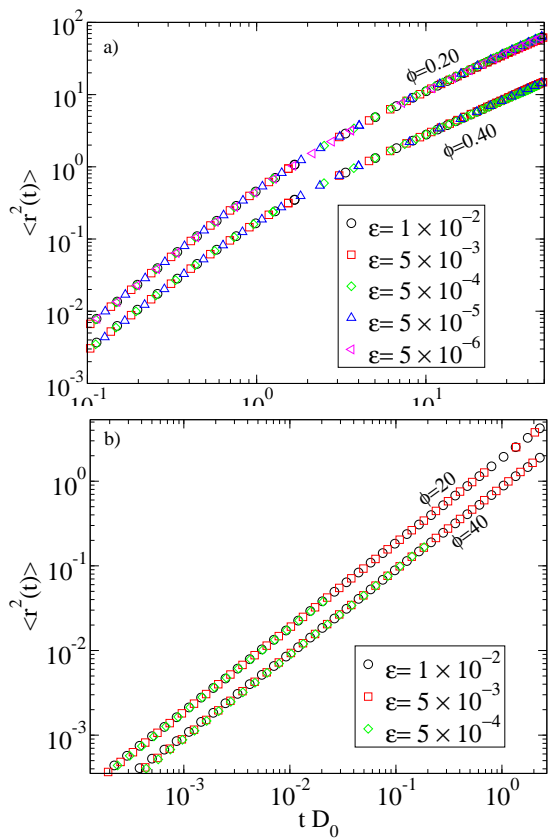

\includegraphics[width=.4\textwidth]{fig3a}
\includegraphics[width=.4\textwidth]{fig3b}
\caption{Mean square displacement for different $\epsilon$ values at the same
$B_2^*=-0.495$ for ND (a) and BD (b) for $\phi=0.20$ and $\phi=0.40$. $D_0 \equiv v_{th} \sigma_{B}$ for ND and 
the bare monomer diffusion constant for BD.
 }
\label{fig:msd}
\end{figure}

Fig.~\ref{fig:msd} shows $<r^2(t)>$ for both ND and BD. 
Data are reported as a function of $t D_0$ to account for the trivial
differences in $v_{th}$  for ND and in the bare self-diffusion
coefficient $D_0$  for BD.  For both $\phi$ and both microscopic
dynamics, $<r^2(t)>$ is independent on the range of the attractive
potential, when the comparison is done at constant $B_2^*$. In other
words, the only difference in the dynamics is accounted by the trivial
microscopic $D_0$ scaling factor.  This implies that the
isodiffusivity curves calculated for the $\epsilon=0.01$ case, when
reported in a $B_2^*-\phi$ plane, describe the entire class of SW
potentials with range shorter than $\epsilon=0.01$.

\begin{figure}[tbh]
\begin{center}
\includegraphics[width=.4\textwidth]{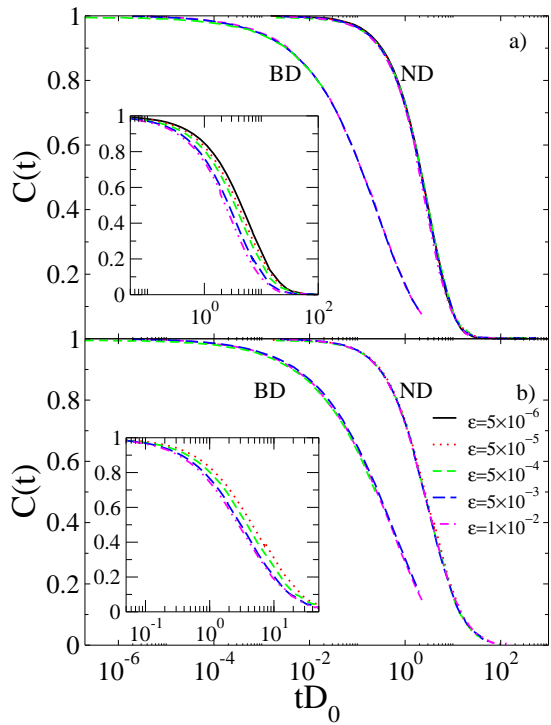}  
\end{center}
\caption{Bond correlation function $C$ as a function of $tD_0$ for different
 relative well width $\epsilon$ and $T$ but all at
 $B^{*}_2=-0.405$. Data for both ND and BD are reported.  (a)
 $\phi=0.20$; (b) $\phi=0.40$. The insets show, for the case of ND,
 $C(t)$ vs $t$.}
\label{fig3}
\end{figure}

Fig.~\ref{fig3} shows $C(t)$ as a function of $t D_0$ for different
pairs ($\epsilon$,$T$) at fixed $B_2^*$.  Since $B_2^*$ is constant by
construction, the average number of bonds in the system is the same
for all investigated ($\epsilon$,$T$) pairs.  In agreement with the
data shown in Fig.~\ref{fig:msd}, all $C(t)$
collapse onto the same curve both for ND and BD.
This suggests that, in $tD_0$ units, the probability of breaking a
bond does not change along constant $B_2^*$ paths.  It is worth
stressing that, while the collapse is observed for both type of
microscopic dynamics, the shape of the ND and BD correlation functions
differs. In ND, $C(t)$ is to a good approximation
exponential while in BD it is stretched, with a stretching
exponent $\approx 0.5$.  The same considerations hold for $\phi=0.40$ (Fig.~\ref{fig3}b).
The fact that the decay of $C(t)$ is still strongly affected by the microscopic
dynamics, implies that MCT can not be applied at these
$\phi$.

In general, assuming that bond breaking is an activated process,
the bond breaking probability can be expressed as a product of a
frequency of bond breaking attempts  $\omega$ times $e^{-\beta
u_0}$, which express the probability of overcoming the barrier. In the
case of ND, $\omega^{-1}$ is proportional to the time requested to
travel a distance of the order of $\Delta$, and $\omega^{-1} \sim
\Delta/v_{th}$. Hence, the bond lifetime, apart from the thermal
contribution $v_{th}$ absorbed in $D_0$, is controlled by the product
$\Delta e^{\beta u_0}$, the same quantity controlling the value of
$B_2^*$ at small $\epsilon$ and $T$.
\\ 
Results presented in Figs.~\ref{fig:msd}-\ref{fig3} suggest that, for
small $\epsilon$, the value of $B_2^*$ characterizes not only
thermodynamics, but also dynamics. In other words, at a given value of
$B_2^*$ it corresponds a family of systems with different $T$ and
$\epsilon$, including the limiting case of Baxter, that posses the
same static and dynamic properties.  According to the present results,
the apparent long bond lifetime characteristic of the Baxter
model~\cite{Miller2003} is only induced by the extremely small thermal
velocity associated to the vanishing of T implicit in the limit
$\epsilon \rightarrow 0$ at fixed $B_2^*$. 
\\
The similar scaling of dynamics and thermodynamics has important
consequences for understanding gel formation in short-range attractive
colloidal dispersions. The isodiffusivity lines reported in
Fig.\ref{fig:1} describe not only the case $\epsilon=0.01$ for which
they have been calculated but also the dynamics of all shorter ranged
potentials, down to the Baxter limit, at least up to the tested
$\phi=0.4$ value. This has a profound consequence for the two
scenarios discussed in Fig.\ref{fig:sketch}, since it proves that in
short-ranged potentials the glass line always meet the liquid-gas line
on its right side. In this class of potentials, disordered arrested
states at low $\phi$ can only be created under out of equilibrium
conditions, requiring a preliminary separation into colloid rich
(liquid) and colloid poor (gas) phases followed by an attractive-glass
dynamic arrest in the denser regions.\\
Authors thank M.~Bruner for help in the preparation of Fig.~1. Support
from MIUR-FIRB and Cofin and Training Network of the Marie-Curie
Programmme of the EU (MRT-CT-2003-504712) is acknowledged.

%
%
%
%
%
%
\bibliography{add,tesi}
\bibliographystyle{apsrev}


\end{document}